\documentclass[pra,twocolumn,superscriptaddress,showpacs,floatfix]{revtex4}

\usepackage{graphics}

\usepackage{bm}
\usepackage{latexsym}
\usepackage{amsmath}
\usepackage{amsfonts}
\usepackage{amssymb}

\newcommand{\ket}[1]{\mbox{$|#1\rangle$}}

\def\identity{\leavevmode\hbox{\small1\kern-3.2pt\normalsize1}}%

\begin{document}

\title{Decoherence can be useful in quantum walks}

\author{Viv Kendon}
\email{Viv.Kendon@ic.ac.uk}
\author{Ben Tregenna}
\affiliation{Optics Section, Blackett Laboratory, Imperial College, London, SW7 2BW, UK.}
\date{September 11, 2002, revised December 14, 2002}

\begin{abstract}
We present a study of the effects of decoherence 
in the operation of a discrete quantum walk on a line,
cycle and hypercube.
We find high sensitivity to decoherence, increasing with the number
of steps in the walk, as the particle is becoming more delocalized
with each step.
However, the effect of a small amount of decoherence 
is to enhance the properties of the quantum walk that are desirable for
the development of quantum algorithms.
Specifically, we observe a highly uniform distribution 
on the line, a very fast mixing time on the cycle, and
more reliable hitting times across the hypercube.
\end{abstract}



\maketitle

\section{Overview}

There is currently great interest within the quantum information community
in quantum versions of random walks, because of the possibility they
may produce new, powerful types of quantum algorithms. 
All the quantum algorithms known until very recently are essentially based on
application of the quantum fourier transform, 
including Grover's search \cite{grover96a} 
and Shor's factoring \cite{shor97a}, which provide quadratic and exponential
speed up respectively over the best known classicl algorithms.
Other types of quantum computation, such as quantum adiabatic computation
\cite{farhi00a}, have not yet been shown
to provide exponential speed up over classical methods.
Some of the most powerful known classical algorithms are based
on classical random walks,
so it is a natural question to ask whether
there are quantum counterparts that can do even better.
For example, a random walk on a general graph
can be used to address hard problems such as
approximating the permanent, $k$-SAT and graph connectivity 
\cite{jerrum01a,schoning99a,motwani95}.

Before attempting to create quantum algorithms from quantum walks,
it is first useful to study their properties and dynamics
on simpler structures.
Several quantum analogues of a
classical random walk on discrete lattices or graphs have been proposed.
These include discrete time walks both with and
without a quantum coin \cite{aharonov92a,watrous01a},
and a continuous time walk \cite{farhi98a}.
The relationship between the continuous time quantum walk and discrete
time quantum walks is not fully understood.
For the cases where they have been studied on the same graph, they give
essentially the same results, e.~g.,~\cite{moore01a}.
Quantum walks on the infinite line, the cycle, and the hypercube
have all been solved analytically \cite{aharonov00a,ambainis01a,moore01a},
and some bounds are known for more general graphs \cite{aharonov00a}.
On the line and cycle, for most quantities of interest,
such as the standard deviation and the mixing time,
there is a quadratic speed up over the classical walk.
Kempe \cite{kempe02a} recently proved that the hitting
time to the opposite corner of a hypercube shows an exponential speed up
(a possibility also found numerically by Yamasaki et al.~\cite{yamasaki02a}).
Making an algorithm out of a quantum walk requires significant further
work.  Two have been proposed very recently, Childs et al.~\cite{childs02a}
use a continuous time quantum walk to traverse a special graph exponentially
faster than a classical algorithm, and Shenvi et al.~\cite{shenvi02a} 
prove that a coined quantum walk can match Grover's algorithm searching an
unstructured database.

Quantum walks themselves (both discrete and continuous time)
can be implemented efficiently on a quantum computer
\cite{aharonov02a,watrous02a,kempe02b}, i.~e., it is not necessary to provide
a physical implementation of a quantum walk to base an algorithm on them.
Nonetheless, quantum walks are interesting in their own right as physical
systems in which a precise level of coherent control can be demonstrated.
Several direct implementations of discrete walks have been proposed,
all with a quantum coin: 
a walk in the vibration modes of a trapped ion \cite{travaglione01a},
in the phase of the field in a cavity containing an atom \cite{sanders02a},
and with an atom hopping between traps in an optical lattice \cite{dur02a}.
Other than for these physical implementations, and recent work on
coin decoherence by Brun et al.~\cite{brun02a,brun02b,brun02c},
the effect of decoherence in quantum walks has not
previously been studied in any detail.

The key observation of this paper is that, in small doses, rather than
degrading the quantum features, decoherence in a coined quantum walk
can enhance the desirable quantum speed up, even though
overall, quantum walks (consisting as they do of extremely delocalized
quantum particles) are highly sensitive to the effects of decoherence.
This is very encouraging for the prospects of using quantum
walks as the basis of powerful quantum algorithms.

The paper is organized as follows.  First we review one of the simplest
examples of a quantum walk, the coined walk on a discrete line, and describe
the properties of the perfect quantum walk.  Then we
present our results showing the effects of decoherence
in the quantum walk on a line, on a cycle, on a hypercube, and on the
``glued trees'' graph of ref.~\cite{childs02a}.

\section{Coined quantum walk}

We consider only coined quantum walks on discrete lattices in this paper.
Since the classical random walk requires a source of randomness
(coin-toss) in the dynamics, introducing a quantum coin is a 
natural way to proceed.  For the walk on an infinite line,
the total Hilbert space is
$H = {\cal C}^{2}\otimes{\cal H}^{\infty}$, where ${\cal H}^{\infty}$
has supposrt on $x \in \mathbb{Z}$.  We label the coin states
by $\{\ket{-\!1},\ket{+\!1}\}$ for ``move left'' and ``move right''
respectively, and those of the particle by $\ket{x}$
for position on the line.
We write the tensor product states as $\ket{a,x}$, where $a \in \{\pm\!1\}$ is
the state of the coin.
The unitary operator describing a single step of the walk is 
\begin{equation}
U=S\cdot(H\otimes \identity)
\end{equation}
where $S$ is the conditional shift operator
$S\ket{a,x}=\ket{a,x+a}$,
and H is the Hadamard operator
\begin{equation}
(H\otimes \identity)\ket{a,x}=(a\ket{a,x}+\ket{-\!a,x})/\sqrt{2},
\label{eq:Hdef}
\end{equation}
acting on the coin states only, it plays the role of a ``coin toss''.
It can be shown for unbiased quantum walks on a line 
in which the ``coin toss'' prepares an
equal superposition of states, all such operators are essentially
equivalent to the Hadamard \cite{ambainis01a,bach02a}.
In higher dimensions, when the particle
has more than two choices of direction at each step, the choice of
unbiased coin-toss operator becomes correspondingly
richer \cite{mackay01a,tregenna03a}.
Unlike the classical case, where each coin toss is independent of the
previous coin tosses, in the quantum case, unitarity of the evolution
and hence reversibility implies the initial state of the coin 
has observable consequences at \textit{all} later times.
Specifically, in the following (unless stated otherwise) we will
choose the initial state to be $(\ket{+\!1}+i\ket{-\!1})/\sqrt{2}$ which
results in a symmetric probability distribution \cite{ambainis01a,bach02a,tregenna03a}.
The dynamics of the quantum walk thus consist of repeated application
of the operator $U$ to the particle and coin, resulting in a spreading out
on the line, with interference causing the quantum speed up.

\begin{figure}
    \begin{minipage}{\columnwidth}
            \begin{center}
            \resizebox{\columnwidth}{!}{\includegraphics{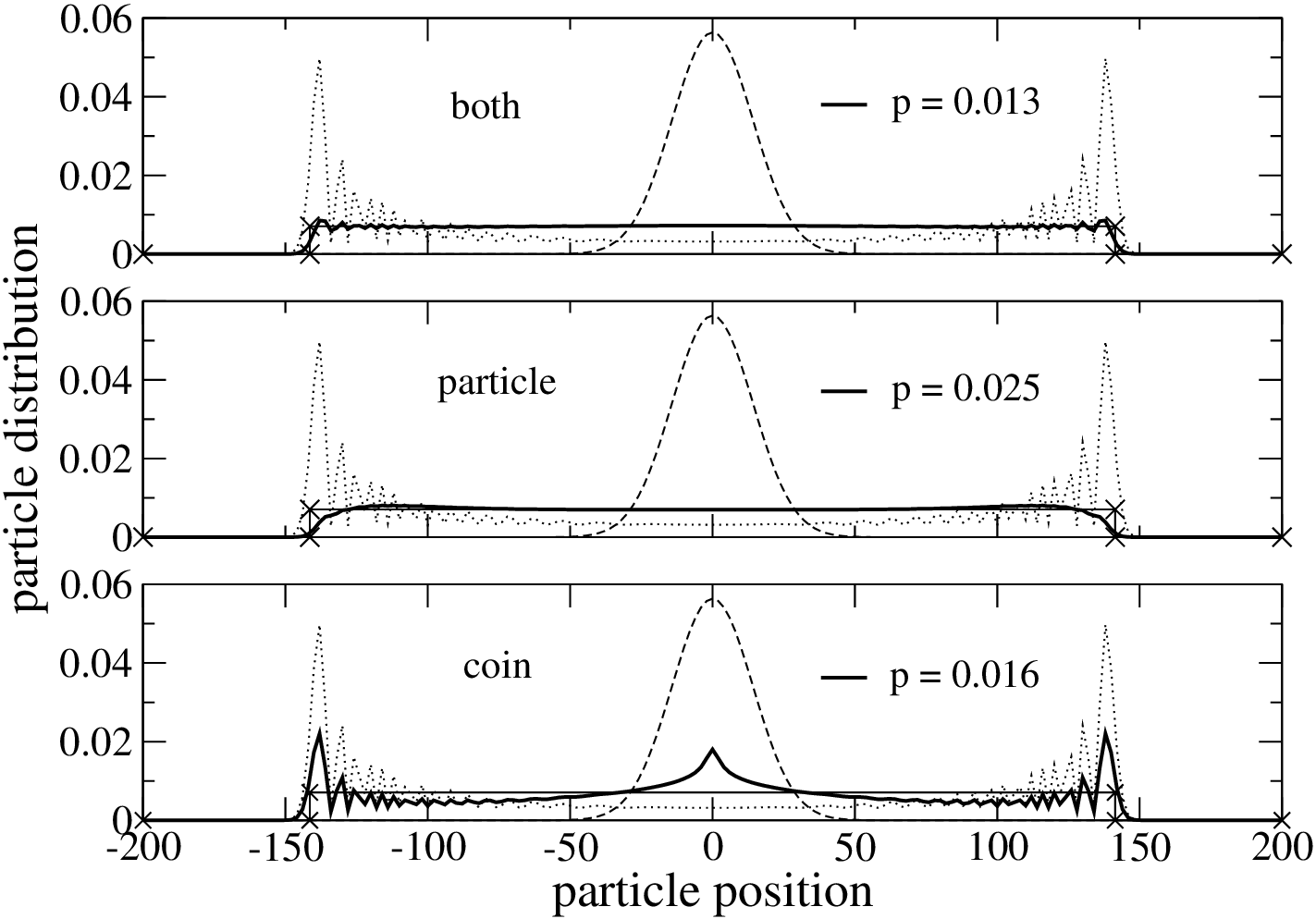}}
            \end{center}
            \caption{Distribution of the particle position 
		for a quantum walk on a line after $T=200$ time steps.
		Pure quantum (dotted), fully classical (dashed),
		and decoherence at rate shown
		on part of system indicated by key (solid).
		Uniform distribution between
		$-T/\sqrt{2} \le x \le T/\sqrt{2}$ (crosses) also shown.}
            \label{fig:200pdf}
    \end{minipage}
\end{figure}
The quantum walk on a line has been solved exactly \cite{ambainis01a}
using both real space (path counting) and Fourier space methods.
The solutions are complicated, mainly due to the
``parity'' property, i.~e., the solutions must have support
only on even(odd)-numbered lattice sites at even(odd) times.
The shape of the probability distribution for the particle position
consists of a nearly flat region around the center with the same width as
the classical binomial distribution, and oscillating peaks out
towards $x = \pm T/\sqrt{2}$. 
Both quantum and classical distributions are shown in fig.~\ref{fig:200pdf},
calculated numerically for $T=200$.
The moments can be calculated, for a walk starting at the origin,
$\langle|x|\rangle = T/2$ and
$\langle x^2\rangle = (1-1/\sqrt{2})T^2 = \sigma^2(T)$.
The standard deviation (from the origin) $\sigma(T)$
is thus linear in $T$, in contrast to $\sqrt{T}$ for the classical walk.

\section{Decoherence in a quantum walk on a line}

\begin{figure}
    \begin{minipage}{\columnwidth}
            \begin{center}
            \resizebox{\columnwidth}{!}{\includegraphics{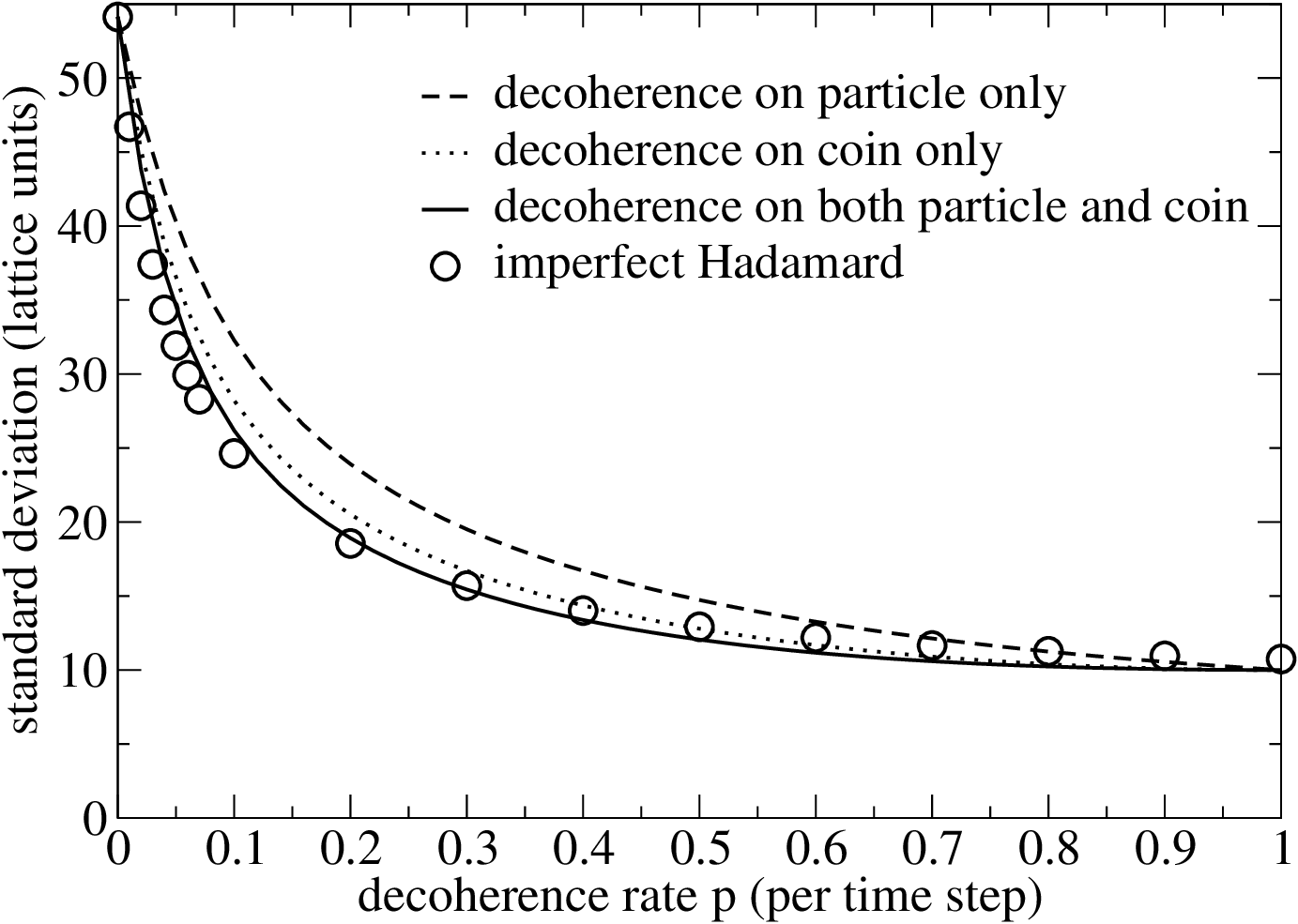}}
            \end{center}
            \caption{Standard deviation $\sigma_p(T)$ of the particle on
		a line for different models of decoherence, for $T= 100$
		time steps.}
            \label{fig:dec100-all}
    \end{minipage}
\end{figure}

In order to model decoherence in this system, we write the
coin-particle dynamics in terms of a density matrix $\rho(t)$ 
that evolves according to
\begin{eqnarray}
\rho(t+1) &=& (1-p)U\cdot\rho(t)\cdot U^{\dag}\nonumber\\
          &+& p\sum_i\mathbb{P}_i\cdot U\cdot\rho(t)\cdot U^{\dag}\cdot\mathbb{P}_i^{\dag}.
\label{eq:decdyn}
\end{eqnarray}
Here $\mathbb{P}_i$ is a projection that represents the action of the
decoherence and $p$ is the probability of a decoherence event
happening per time step.  We took equation (\ref{eq:decdyn})
and evolved it numerically for various choices of $\mathbb{P}_i$.
Motivated by the likely form of experimental errors,
we also modelled an imperfect Hadamard 
by applying a Gaussian spread of standard deviation $\sqrt{p}\pi/4$
about the perfect value of $\pi/2$ implicit eq.~(\ref{eq:Hdef}),
compare \cite{mackay01a}.
An imperfect shift on the particle has been studied in \cite{dur02a}.
In each case we find the same general form for the decay of 
$\sigma_p(T)$ from the quantum to the classical value,
with small differences in the rates, as shown in fig.~\ref{fig:dec100-all}.
The slope of $\sigma_p(T)$ is finite as $p\rightarrow 0$ and zero
at $p = 1$.  We calculated $\sigma_p(T)$ analytically
(details in ref.~\cite{tregenna02a})
for $pT \ll 1$ and $T \gg 1$ for the case where $\mathbb{P}_i$ is the
projector onto the prefered basis $\{\ket{a,x}\}$ (decohering both
particle and coin),
\begin{equation}
\sigma_p(T) \simeq \sigma(T)\left\{1-\frac{pT}{6\sqrt{2}} + O(p) \right\}.
\end{equation}
This compares well with simulation data, once a second order correction
for $\sigma(T) = (1-1/\sqrt{2})^{1/2}(T-1/T)$ is taken into account.
The first order dependence is thus proportional to $pT$, 
so the sensitivity to decoherence grows linearly
in $T$ for a given decoherence rate $p$. 

The quantum walk on infinite regular lattices of higher dimension shows a
similar decoherence profile, as might be expected, since
the standard deviation of a classical random walk is $\sqrt{T}$, independent
of dimension.

\section{Distribution shape on a line}

\begin{figure}
    \begin{minipage}{\columnwidth}
            \begin{center}
            \resizebox{\columnwidth}{!}{\includegraphics{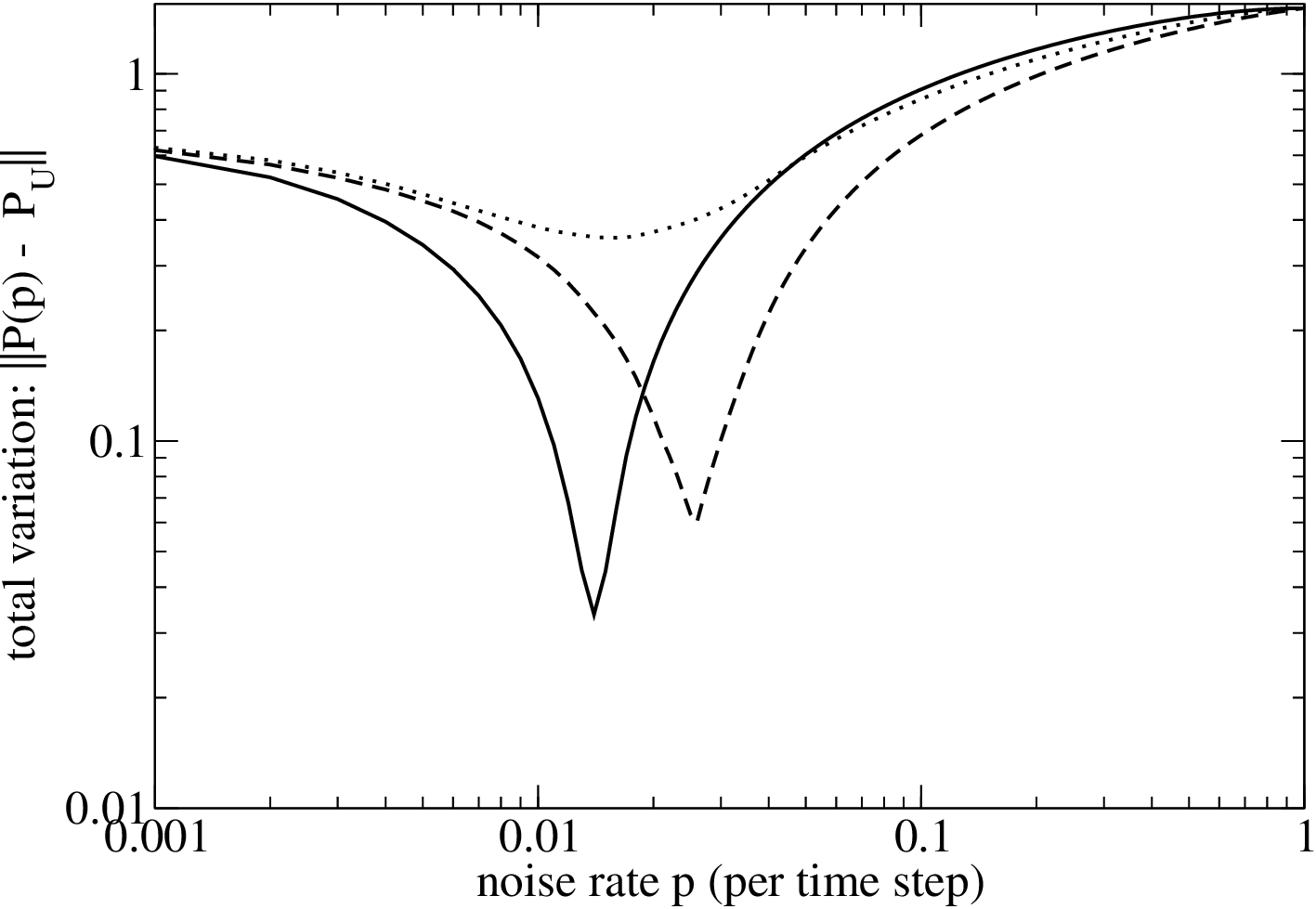}}
            \end{center}
            \caption{The total variational distance
                of the particle position distribution from the uniform
                distribution for T=200 and decoherence on coin (dotted),
		particle (dashed) or both (solid).}
            \label{fig:200flat}
    \end{minipage}
\end{figure}

The transition from quantum to classical in the standard deviation of the
particle position shown in fig.~\ref{fig:dec100-all} is qualitatively
the same for all types of decoherence examined.
However, there are interesting differences in the shape of the
distribution of the particle position.  
The decoherence rate that gives the closest to uniform distribution has been
selected and plotted in fig.~\ref{fig:200pdf},
along with the pure quantum and classical distributions
for comparison.  When the particle position is subject
to decoherence that tends to localize the particle in the standard basis,
this produces a highly uniform distribution between $\pm T/\sqrt{2}$ for a
particular choice of $p$.  The optimal decoherence rate $p_u$ can be obtained
by calculating the total variational distance between the actual and
uniform distributions,
\begin{eqnarray}
\nu(p,T) &\equiv& ||P(x,p,T) - P_u(T)||_{\text{tv}}\nonumber\\
	 &\equiv& \sum_x|P(x,p,T) - P_u(T)|,
\label{eq:tvd}
\end{eqnarray}
where $P(x,p,T)$ is the probability of finding the particle at position $x$
after $T$ time steps, regardless of coin state,
and $P_u(T) = \sqrt{2}/T$ for $-T/\sqrt{2} \le x \le T/\sqrt{2}$
and zero otherwise.  Figure \ref{fig:200flat} shows $\nu(p,T)$ for $T=200$
with decoherence applied to the coin, particle and both at once.
Decoherence on both particle and coin produces the best uniform
distribution, but at a cost of a lower decoherence rate, and sharper minimum,
i.~e., greater sensitivity to the value of $p$.
Walks with decoherence only on the particle can tolerate more
variation in the exact decoherence rate, while not
achieving such a good uniform distribution.  
The optimum decoherence rate depends on the number of steps in the walk,
we determined numerically that $p_uT \simeq 2.6$ for decoherence on both
and $p_uT \simeq 5$ for decoherence on the particle only.
These differences in the quality of the uniform distribution are 
independent of $p$ and $T$, and provide an order of magnitude
(0.6 down to 0.06) improvement in $\nu$ over the pure quantum value.
Decoherence just on the coin doesn't enhance the uniformity of the
distribution, as fig.~\ref{fig:200pdf} shows, there is a cusp at $x=0$.
However, for finite $T$,
there is still a useful window within which even coin only
decoherence does not significantly degrade the linear spreading of the walk.

\section{Decoherence on a cycle}

If, instead of an infinite line, the quantum particle is allowed to walk
on a cycle of size $N$ points, the appropriate quantity to measure the
progress of the walk is the mixing time.  We immediately have to modify
the classical definition, because, unlike the classical random walk
which mixes to a uniform distribution in the the long time limit,
a unitary process such as that evolving the quantum walk
does not mix to any limit at large times.  We can instead define a time
averaged particle distribution
\begin{equation}
\overline{P(x,p,T)}=\frac{1}{T}\sum_{t=0}^{T-1} P(x,p,t)
\label{eq:meanP}
\end{equation}
which does always mix for large enough times $T$ \cite{aharonov00a}.
It is easy to sample from the distribution $\overline{P(x,p,T)}$:
run the walk for some randomly chosen number of steps $0<t<T$ and
measure the particle position at that time $t$.
The mixing time is then defined as
\begin{equation}
M_{\epsilon}=\min\left\{T|\forall t>T: ||\overline{P(x,p,t)}-P_u||_{\text{tv}}<\epsilon\right\}
\label{eq:mixdef}
\end{equation}
where $P_u$ is the limiting (uniform) distribution over the cycle.
The mixing time quantifies how long it takes for
the time-averaged probability distribution of the particle position to 
reach its limiting value within a margin of small parameter $\epsilon>0$.

The walk on a cycle is the simplest example of a walk on the Cayley graph of an Abelian group, and was historically the first to be treated analytically
\cite{aharonov00a}.
The dynamics of the walk on the cycle are the same as for the walk on a 
line, with the particle position taken mod($N$).
Aharonov et~al.~\cite{aharonov00a} proved an upper bound for the mixing time
of $O(N\log N)$.
The limiting distribution thus obtained depends on the choice of coin
operator \cite{tregenna03a}, in sharp contrast to the classical walk, which
always mixes to a uniform distribution.
For the Hadamard coin used here, the odd-$N$
cycle mixes to the uniform distribution, but the even-$N$ cycle does not.

\begin{figure}
    \begin{minipage}{\columnwidth}
            \begin{center}
            \resizebox{\columnwidth}{!}{\includegraphics{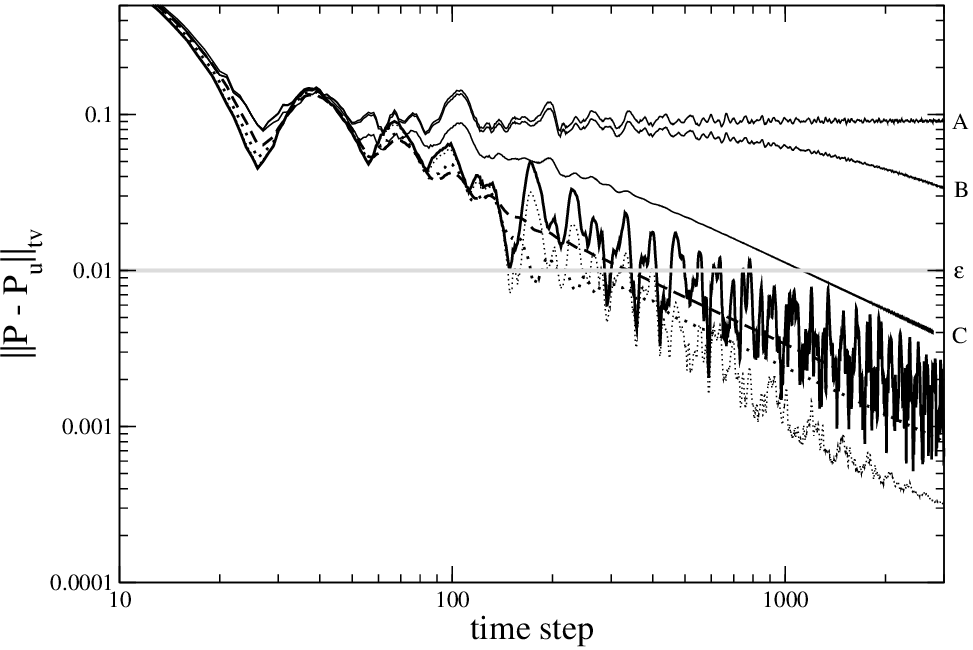}}
            \end{center}
            \caption{Difference between the time averaged probability
		distribution $\overline{P(x,p,T)}$ and the uniform distribution
		expressed as the total variational distance,
		eqs.~(\ref{eq:tvd})--(\ref{eq:meanP}) for the case with
		decoherence on both coin and particle.
		Both axes logarithmic.  The value of $\epsilon$ used in
		fig.~\ref{fig:c30c29} is shown as a horizontal line.
		Top three solid lines labeled on right are for $N=22$ with
		$p=0$ (A), $p=0.001$ (B), $p=0.02$ (C).
		Lower four lines are for $N=21$ with $p=0$ (solid),
		$p=0.002$ (dotted), $p=0.01$ (short dashed),
		$p=0.02$ (long dashed).}
            \label{fig:epsilon}
    \end{minipage}
\end{figure}

\begin{figure}
    \begin{minipage}{\columnwidth}
            \begin{center}
            \resizebox{\columnwidth}{!}{\includegraphics{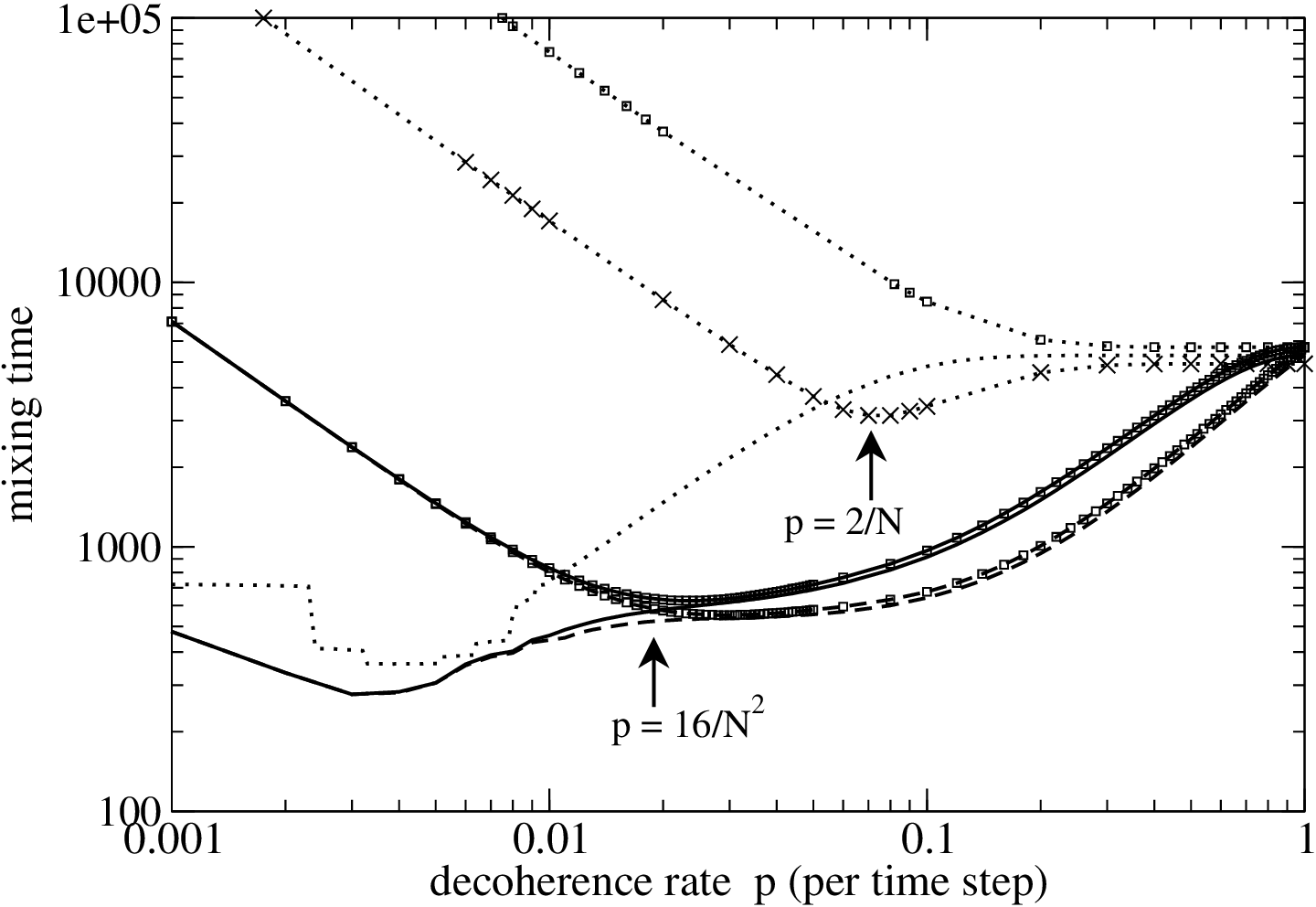}}
            \end{center}
            \caption{Numerical data for mixing times on cycles of size
		$N=29$ and $N=30$ ($\square$), for coin (dotted), particle
		(dashed) and both (solid) subject to decoherence,
		using $\epsilon=0.01$. Also $N=28$ ($\times$) for coin.
		Both axes logarithmic.}
            \label{fig:c30c29}
    \end{minipage}
\end{figure}

We numerically evaluated the mixing times for walks on cycles of sizes
up to $N\simeq 80$, both for pure states, and in the presence of the same
types of decoherence as described in the previous section
for the walk on a line.
For odd-$N$ cycles with no decoherence,
we find that $M_{\epsilon} \sim N/\epsilon$ 
as compared to the upper bound of $M_{\epsilon} \sim N \log N/\epsilon$
in ref.~\cite{aharonov00a}.  While we believe that the linear scaling with
$N$ is the correct result, obtaining a tighter bound analytically
is a tough task, because the time averaged probability distribution
$\overline{P(x,p,T)}$ is a rather fluctuating quantity, especially for $p=0$,
as is illustrated in fig.~\ref{fig:epsilon}.  In this figure, the quantity
$||\overline{P(x,p,t)}-P_u||_{\text{tv}}$ from eq.~(\ref{eq:mixdef}) is
plotted against time.  The time at which
the curves last cross the horizontal line at $\epsilon=0.01$ is the mixing
time as plotted in fig.~\ref{fig:c30c29}.
A different choice of $\epsilon$ thus causes a jump in the value of
$M_{\epsilon}$ if it happens to touch the next peak in
$||\overline{P(x,p,t)}-P_u||_{\text{tv}}$.

\subsection{Even-$N$ cycles with decoherence}

Under the action of a small amount of decoherence, the mixing time becomes
shorter for all cases, typical results are shown in fig.~\ref{fig:c30c29}.
Also, decoherence causes the even-$N$ cycle to mix to the
uniform distribution.
The asymptotes in fig.~\ref{fig:c30c29} for $N$ even and decoherence on
the coin only, for $p < 2/N$, are well fitted by
$\epsilon p M_{\epsilon} \simeq N/4$ for $N$ divisible by 2, and
$\epsilon p M_{\epsilon} \simeq N/16$ for $N$ divisible by 4.
For larger $p$, the mixing time tends to the classical value
of $N^2/16\epsilon$ [note this is not $\log(1/\epsilon)$ because
we are calculating the average mixing time, eq.~(\ref{eq:mixdef})].
Although for $N$ divisible by 4, the (coin-decohered)
mixing time shows a minimum below the
classical value at $p \simeq 2/N$, this mixing time is $\gtrsim N^2/32\epsilon$,
i.~e.,~still quadratic in $N$.
Thus, although noise on the coin causes the even-$N$ cycle to mix
to the uniform distribution, it does not produce a significant speed up over
the classical random walk.

For decoherence on the particle position, with $p < 16/N^2$,
$\epsilon p M_{\epsilon} \simeq 1/(N/2-1)$ for $N$ divisible by 2, and
$\epsilon p M_{\epsilon} \simeq 1/(N/4+3)$ for $N$ divisible by 4.
At $p \simeq 16/N^2$, there is a minimum in the mixing time at a value
roughly equal to the $(N\pm 1)$-cycle pure quantum mixing time,
$M^{(\text{min})}_{\epsilon} \sim \alpha N/\epsilon$ (with $\alpha$ a
constant of order unity).
The top three lines in fig.~\ref{fig:c30c29} show how decoherence
pulls an even-$N$ cycle down to mix
to the uniform distribution at the same rate as the neighboring odd-$N$ cycle.
Decoherence on the particle position thus causes the even-$N$ cycle to
mix to uniform in linear time for a suitable choice of decoherence
rate $p^{(\text{min})} \sim 16/N^2$, independent of $\epsilon$.

\subsection{Odd-$N$ cycles with decoherence}

For all types of decoherence, the odd-$N$ cycle shows a minimum mixing
time at a position somewhat earlier than the even-$N$ cycle,
roughly $p = 2/N^2$,
but because of the oscillatory nature of $\overline{P(x,p,T)}$,
the exact behavior is not a smooth function of $p$ or $\epsilon$.  
As decoherence on the particle (or both) increases, 
the oscillations in $\overline{P(x,p,T)}$ are damped out.
The lower set of lines on fig.~\ref{fig:epsilon} shows how both these features
affect the mixing time.
At $p \simeq 16/N^2$, the mixing time passes smoothly
through an inflexion and from then on behaves
in a quantitatively similar manner to the adjacent-sized even-$N$ cycles,
including scaling as $M^{(\text{min})}_{\epsilon} \sim \alpha N/\epsilon$
at the inflexion.  Thus for $0 \le p \lesssim 16/N^2$ there is a region
where the mixing time stays linear in $N$.
Our overall conclusion is thus the same as for the walk on a line, there
is a useful window within which decoherence enhances rather than degrades
the quantum features of the walk.

\section{Hypercube decoherence}

The hypercube (boolean $N$-cube) is also the Cayley graph of a group,
and provides a step on the way to quantum walks on more complex structures.
Since there are $N$ edges joining at each vertex, we need an $N$-dimensional
coin to choose between the possible paths at each step.  This opens up a 
correspondingly larger range of possible unitary operations to use for
the coin toss, but a sensible choice is one which respects the symmetry of
the underlying graph, and is as far from the identity as possible.  For
the hypercube, this is the Grover operator, whose elements expressed as an
$N\times N$ matrix are defined as $G_{jk}=2/N-\delta_{jk}$.
We also choose a symmetric starting state for the coin with equal
weights for all possible directions from the chosen starting node.
The high degree of symmetry in the hypercube allows the quantum walk
to be mapped to a walk on a line with a different coin toss operation
at each point \cite{moore01a}.
While the mixing time for a quantum walk on a hypercube
is somewhat worse than a classical random walk, Kempe \cite{kempe02a}
proved that the hitting time to the opposite corner
is polynomial, an exponential speed up over the classical walk.

Kempe discusses two types of hitting times, one-shot,
where a measurement is made after a pre-determined number of steps,
and concurrent, where the desired location is monitored
continuously to see if the particle has arrived.
In each case, the key parameter is the probability $P_h$ of
finding the particle at the chosen location.
\begin{figure}
    \begin{minipage}{\columnwidth}
            \begin{center}
            \resizebox{\columnwidth}{!}{\includegraphics{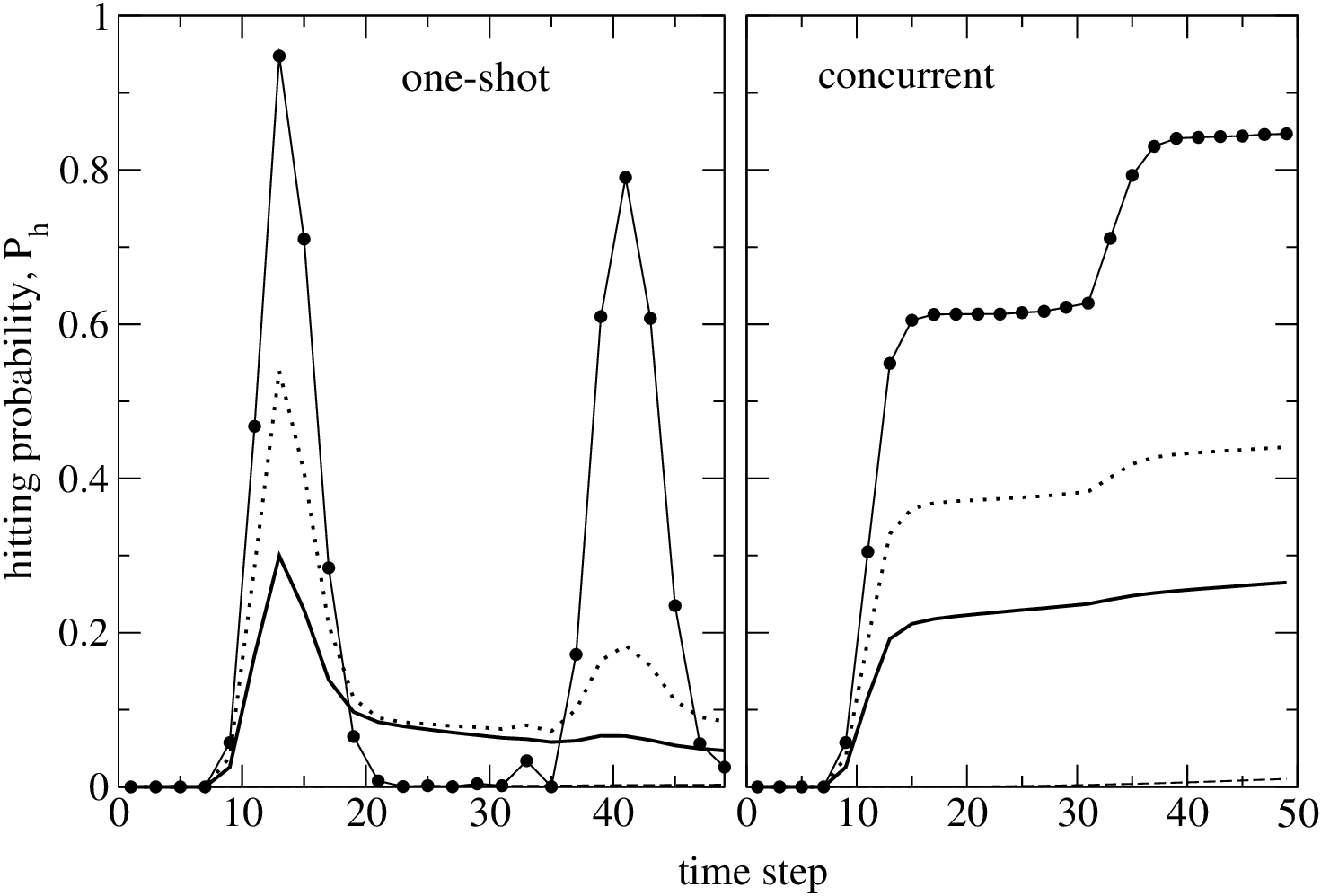}}
            \end{center}
            \caption{Hitting probability on a 9--dimensional hypercube
		for one-shot (left) and concurrent (right),
		perfect walk (circles),
		with $p = 0.05$ (dotted), $p = 0.1 \simeq 1/9$ (solid).
		Classical hitting probability barely visible (dashed).}
            \label{fig:9hit}
    \end{minipage}
\end{figure}
We calculated $P_h$ numerically and found that
all forms of decoherence have a similar effect on $P_h$,
see fig.~\ref{fig:9hit}, reducing the peaks and smoothing out the troughs.  
For the one-shot hitting time this is useful, 
raising $P_h$ in the trough to well above the classical value,
so it is no longer necessary to know exactly when to measure.
For $p \lesssim 1/N$, the height of the first peak scales as
$P_h(p) = P_h(0) \exp\{-(N+\alpha)p\}$,
where $0\lesssim\alpha\lesssim 2$ depending on whether coin, particle
or both are subject to decoherence.
An exponential decrease in the presence of decoherence sounds about as
bad as it could reasonably be, and for long times, of course, decoherence
reduces the walk to classical behavior.
However, the hitting times are short, only $\sim N\pi/2$ steps,
and $p\simeq 1/N$ only lowers $P_h$ by a factor of $1/e$.
For algorithmic purposes this is insignificant, only a factor of
order unity and thus still exponentially better than classical.
(Standard amplification techniques can be used to bring the hitting
probability as close to one as desired.)  Note also, that the
size of the graph (measured in number of nodes) is exponential in $N$,
so the decoherence has only a linear effect measured in terms of
the size of the graph.

Continuous monitoring of the target location as in the concurrent
hitting time is already a sort of controlled decoherence, note that
the height of the initial peak at $\sim N\pi/2$ steps is only about $2/3$
that of the one-shot hitting probability.
No extra features are produced by the addition of unselective decoherence,
but there is still a range of $0 < p \lesssim 1/N$ within which
the quantum speed up is preserved.
Note that in both the one-shot and concurrent cases, $p\simeq 1/N$
is a critical damping rate, smoothing out the second peak (shown at
around 40 ($\equiv 3N\pi/2$) steps in fig.~\ref{fig:9hit}).

\section{``Glued trees'' walk}

The graph used in Childs et al.~\cite{childs02a} is similar to the hypercube
in that it can also be mapped to a walk on a line.
It consists of two identical binary trees of depth $N$, glued at their
branches by a randomly distributed set of edges (two per node so the degree
of the graph is three except for the roots of the trees).
They use a continuous time walk for their algorithm, but a discrete
quantum walk using a three dimensional Grover coin operator has
essentially similar properties \cite{watrous02a,tregenna03a}.
The effects of decoherence are similar to the hypercube.  The peak in
the probability of finding the particle at the exit node occurs after
$2N+1$ steps, and is around $0.6$ for a pure walk.
Decoherence reduces this peak exponentially (in $N$)
while spreading out the range of time steps over which the
probability is significantly (exponentially) larger than the classical value.

\section{Conclusions}

One of the generic ways in which classical
random walks are applied to algorithms is to sample an exponentially
large problem space to estimate statistical properties of the system.
Fast mixing times and uniform sampling are necessary
properties of the random walk process for it to perform efficiently.
This is the sense in which we propose that a small amount of
decoherence in a walk on the line or the cycle is beneficial,
producing more uniform distributions (line)
and faster mixing to a uniform distribution (cycle).

On the hypercube and ``glued trees'' walks, the key quantum
property exploited by the examples in the literature
is the opposite of a uniform distribution, the ability of a quantum walk
to continue its forward march through the graph despite the many possible
``wrong turns'' a classical random walk gets lost in.
For the examples given, it is easy to determine \textit{a priori} the best time
to check for the particle having found the target node.  But for more
complex graphs this may not be easy to calculate, so having a wider window
of opportunity to successfully detect the particle could be an advantage.
Note that the concurrent hitting time is doing exactly this in a more
precise way by monitoring the target node at every time step.
For these type of walks, we don't claim that decoherence gives a major
advantage, only that it is not detrimental for small decoherence rates.

We have also found numerically that the walk on a cycle mixes in
linear time (compared to the upper bound of $O(N\log N)$ proved in
\cite{aharonov00a}), and show why it is hard to prove a tighter bound
analytically because the quantities involved are highly fluctuating.

Overall, these results are
quite promising for the development of further quantum algorithms,
and for the practical implementation of quantum computing.
They are also encouraging for the experimental implementation of
quantum walks as proposed in refs.~\cite{travaglione01a,sanders02a,dur02a}.
For the modest number of steps in the proposals, our results suggest
a very reasonable range in which quantum effects can be observed experimentally.
Finally, it is also an intriguingly counter-intuitive result in its own right,
worthy of further study in the context of the transition from
quantum to classical.

\begin{acknowledgments}
We thank Peter Knight, Richard Maile, Will Flanagan, Julia Kempe,
Todd Brun, Hilary Carteret, John Watrous and Dorit Aharonov
for useful discussions.
Work funded by the UK Engineering and Physical Sciences
Research Council.
\end{acknowledgments}

\bibliography{qrw,ent}

\end{document}